\documentclass[twocolumn,aps,prl,showpacs,amsmath]{revtex4}
\usepackage{graphicx}
\usepackage{color}
\usepackage{amsmath}
\usepackage[american]{babel}

\begin{document}

\title{Curvature effect on the interaction between folded graphitic surface and silver clusters}
\author{A. F. Kemper and H-P. Cheng}
\affiliation{Department of Physics and the Quantum Theory Project,
University of Florida, Gainesville, FL 32611, USA}

\author{N. K{\'e}ba{\"i}li, S. Benrezzak, M. Schmidt, A. Masson and C. Br{\'e}chignac}
\affiliation{Laboratoire Aim{\'e} Cotton, CNRS, Univ Paris-Sud 11,
Bat 505, Campus d'Orsay, 91405 Orsay Cedex, France}


\begin{abstract}
Evidence of curvature effects on the interaction and binding of silver clusters on folded graphitic surfaces has been shown from both experiment and theory.
Density Functional Theory (DFT) calculations within the local density and generalized gradient approximations have been performed for the structural relaxation of both Ag and Ag$_2$ on curved surfaces, showing a cross-over from quantum to classical behaviour. Using Lennard-Jones potential to model the interaction between a single cluster and the graphene surface, evidence is found for the curvature effect on the binding of silver nano-particles to folding graphitic surfaces. The theoretical results are compared to SEM and AFM images of samples obtained from pre-formed silver cluster deposition on carboneous substrates exhibiting anisotropic pleat structures.
\end{abstract}
\maketitle

{\em{Introduction.}}\
In the quest for nano-scale science and technology, interaction and diffusion of adatoms, molecules and nanoparticles on surfaces attract enormous interest due to their relevance in the construction and the stability of new nanoarchitectures.
Most studies of adatom diffusion have focused on flat surfaces and it is only since a decade ago that the study of diffusion on deformed surfaces has opened up new perspectives in anisotropic diffusion. It has been shown that the interaction of silver atoms with carbon nanotubes is curvature specific\cite{dshu01}. Both experiments and simulations pointed out that convex surfaces, such as those on the outside of carbon nanotubes, enhance binding of silver atoms. In contrast, our recent experiment on silver cluster deposition on folded graphite shows evidence of repulsive barriers for convex bends \cite{mschmidt08}. \\
On graphite surfaces, metal clusters of up to a few thousand atoms are known to be highly mobile. They aggregate into fractal islands\cite{byoon99,alando06} through an isotropic diffusion of clusters on flat terraces, which has been understood using a Diffusion Limited Aggregation (DLA) model\cite{twitten81}. The islands with anisotropic morphologies obtained on curved surfaces reveal anisotropy in cluster diffusion, which was attributed to surface curvature. In fact, the weak interaction between two adjacent graphene sheets show evidence of pleats, observed via AFM imaging techniques. These pleats destroy the homogeneity of the graphite surface, which should affect cluster mobility and thus change the island patterns\cite{mschmidt08}. Understanding the role of the surface curvature on cluster mobility becomes of fundamental interest for building new architectures. However, nothing is known on the mechanism underlying the diffusion, and in particular the binding interaction between the cluster and folded graphite has not been studied. Two main questions need to be answered to fully understand and predict the final growth pattern from nanoparticle diffusion and aggregation on surfaces. Since atoms and clusters diffuse with opposite behaviours on convex curvature, at which cluster size does the transition from atom to cluster occur? And, for clusters, how does the cluster-surface interaction depend on the surface curvature? \\
In this letter, we show evidence of a curvature effect on the interaction between silver nano-particles and folding graphitic surfaces, both experimentally and theoretically. We show that the islands grown from silver nanocluster deposition on graphite remain fractal on long range undulated surfaces, as it is for flat terraces, whereas those grown on more concave region of the pleats are elongated structures that are located in the valley of the pleat. Several levels of theoretical methods, from quantum mechanics to continuum model, have been used to address various aspects of the problem. Using Density Functional Theory we note that the binding energy of an atom and a dimer on a perfect carbon nanotube behave in opposite ways. For larger clusters, using a Lennard-Jones potential to model the interaction between a single cluster and the graphitic surface, we provide quantitative information on binding energy changes as function of surface curvature. The theoretical results fully explain the experimental observations on cluster deposition on carboneus substrates.

\begin{figure}[ht]
\includegraphics[width=3.2in]{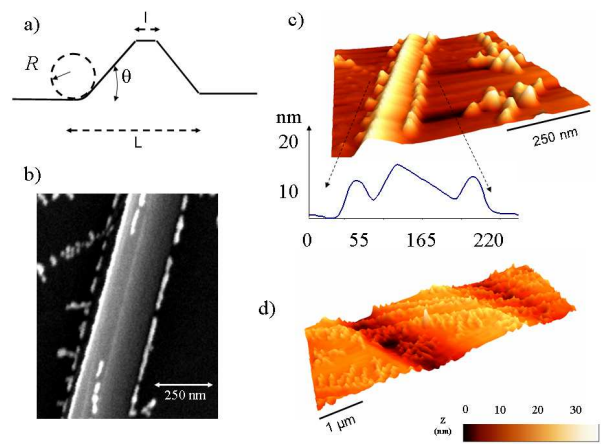}
\caption{(Color online) (a) Schematic drawing of a pleat with the top width l and bottom width L, as well as its concave radius of curvature R and angle $\theta$. (b) and (c) SEM and AFM images of a graphite pleat of 250 nm width after cluster deposition. Notice that islands grow only on the top faces and convex valley. $\theta$ equals 18 and 7 degrees for the concave and convex bends, respectively (note that the pleat is asymmetric and the convex bends were not measured) (d) AFM image of a larger pleat, few $\mu m$ width and comparable height to (c). In this case the angle of curvature is 3 degrees and fractals patterns are not affected by the curvature.}
\label{fig:expt}
\end{figure}

{\em{Experiment.}}\ In our experiments, a distribution of neutral silver clusters with a mean diameter of 3 nm and half-width at half maximum of 0.5 nm is deposited at thermal energy on folded graphite. The low impact energy of 0.05 eV/atom, as compared to the Ag-Ag \cite{mschmidt03} binding energy 1.2 eV, makes the fragmentation of the impinging clusters unlikely. They diffuse on the surface as a whole and grow into islands. The island morphology of the samples is separately analyzed by SEM and tapping AFM. On graphite terraces the clusters aggregate to fractal islands anchored on point defects as described before \cite{byoon99,alando06}. In this paper we focus on cluster deposition on curved graphitic surfaces. The pleats are typically many $\mu m$ long and resemble macroscopic drapery \cite{hhiura94}, even though the graphite is mostly facetted like graphite polyhedral crystals as drawn in Fig. \ref{fig:expt}(a). We focus on three-faced pleats, whose width ranges from 100 nm to a few microns. Atomic force microscopy (AFM) reveals that their height ranges from 10 to 35 nm, and the bend angles $\theta$ of the pleats is quite low -- typically range between 3 and 35 degrees. Although the angles of the pleats are quite well-defined from their AFM profiles, the estimated values of the radius of curvature are less accurate and are limited by the size of the AFM tip -- typically $\sim$10 nm. Figure \ref{fig:expt}(b) and \ref{fig:expt}(c) show typical SEM and AFM images of a graphite pleat of 250 nm width after silver cluster deposition. As described earlier\cite{mschmidt08}, the clusters aggregate to linear islands in the concave bends of the pleat, in which they are trapped, but not demobilized as on point defects or step edges. We reported further, that convexly bent areas act as effective repulsive barriers for the cluster diffusion, which the clusters cannot pass by their thermal diffusion. Such potential barriers are therefore at least 25 meV high (thermal energy). By contrast, the islands grown on slightly curved graphite show isotropic fractal morphology that settle with the shape of the graphitic surface (Fig. \ref{fig:expt}(d)). This indicates that cluster mobility is not affected by graphite curvature of small angle $\theta$ equal to 3 degrees.

{\em{DFT calculation.}}\ 
To compare the behaviour of atoms and clusters on curved surfaces, we start by calculating binding energy of a silver atom and dimer on curved graphene. We used perfect carbon nanotubes (CNTs) to model the curved graphene.
The electronic structure and structural relaxation calculations were performed using Density Functional Theory (DFT)\cite{wkohn65} within the local density and generalized gradient approximations as implemented in the Quantum-ESPRESSO\cite{pwscf} package. In our calculations, plane wave basis sets, PBE exchange-correlation functional\cite{jperdew96} and LDA potential, and RRKJ ultrasoft pseudopotentials\cite{arappe90} have been employed. We used nonlinear core corrections for the Ag atoms, with the 3d state included in the valence. The use of ultrasoft pseudopotentials enabled us to use an energy cut-off of 32 Ry for the plane wave basis, while the density cut-off was taken to be 400 Ry. The Brillouin zone was sampled by $1 \times 8\times 1$ special k-points using the Monkhorst-Pack scheme\cite{hmonkhorst76}, and a Gaussian smearing of width $\sigma=0.001$ Ry was used for electron occupations.\\

\begin{figure}[ht]
\includegraphics[width=3.2in]{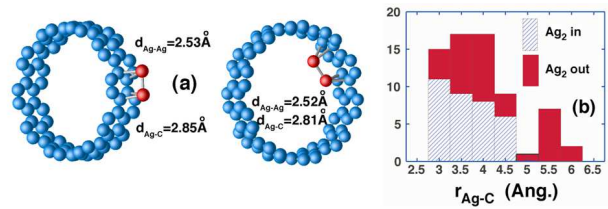}
\caption{(Color online) (a) Adsorption of Ag$_2$ on the outer and inner surfaces of a (10,10) CNT, d$_{\textrm{\textit{Ag-C}}}$ is the average of 4 nearest Ag-C distances. (b) distribution of the first 33 Ag-C distances, the filled box corresponds to Ag$_2$ outside the CNT (left in panel a), and shaded outside the CNT (right in panel a).}
\label{fig:structplot}
\end{figure}

We started with a single Ag atom, comparing the binding energy as calculated by DFT-GGA on the concave and
convex surface. On a (5,5) nanotube, the silver atom is more strongly bound to the convex surface (outside) by 0.20
eV. On a (10,10) CNT, which has lower curvature, the two binding energies are nearly equal.
We then computed Ag$_2$ on a (10,10) CNT (see Fig. \ref{fig:structplot}(a))
and found that the dimer prefers inside site by $\sim$0.3 eV 

In a recent work by Jalkanen et al., it was noted that the van der Waals interaction is a large component of the binding between silver and graphene, and GGA may not be an adequate approximation for the small curvature limit\cite{jjalkanen07}.
To examine this issue, we repeated the above calculations using LDA. Our results show that the Ag atom
energetically prefers the convex surface of the (10,10) CNT by $\sim$0.1 eV. As in the GGA calculations, the dimer
Ag$_2$ is more strongly bound to the concave surface of a (10,10) CNT by $\sim$0.2 eV, in contrast to the single atom.
The unbound atom prefers the convex surface where the stretched C-C bonds allow for some chemical binding. The dimer, on
the other hand, displays classical behavior, i.e. it can be modeled by a pair-wise additive potential.

The concave surface provides a closer area of contact, as illustrated in Figure \ref{fig:structplot}(b),
and thus higher binding.
This means that Ag$_n$ clusters can be modeled classically for n$\ge$2.

For single Ag atoms, quantum-mechanical effects need to be taken into account explicitly.
A further argument for the classical picture for n$\ge$2 comes from the magnitude of the binding energy. To
date, there is no direct experimental data on the desorption energy of a Ag atom or cluster on a graphite surface.
Theoretical calculations give values of binding energy and bond length with a large error bars (a few tenth of eV). We have
performed LDA calculations for a Ag$_{20}$ cluster on a flat graphene
sheet. The binding energy between Ag$_{20}$ and the graphene sheet was 1.20 eV. The clusters are
relatively weakly bound, again confirming the validity of a classical model.

\begin{figure}[ht]
\includegraphics[width=3.2in,clip=true]{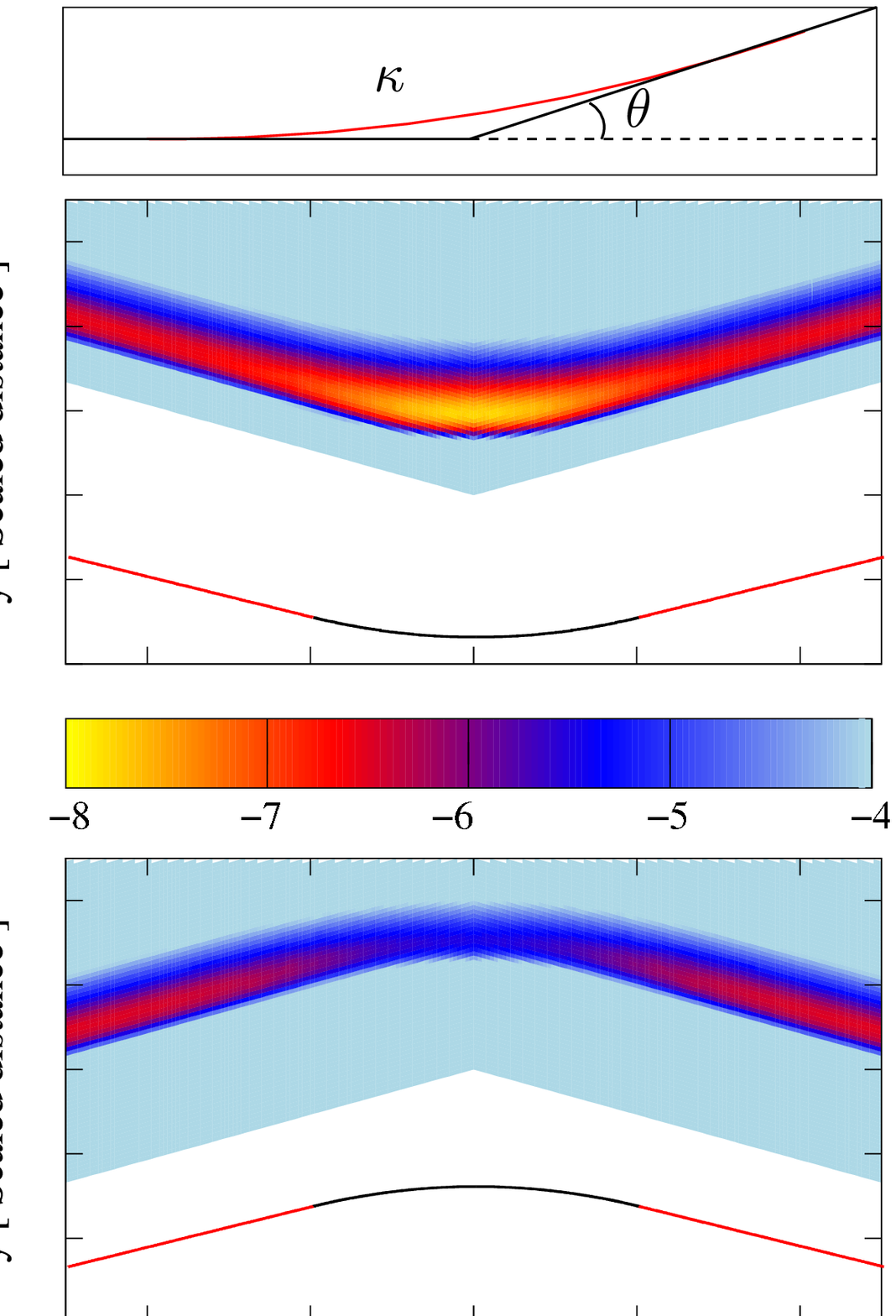}
\caption{(Color online) Top: Diagram of model bend.
Middle: The potential surface, scaled by $4\epsilon$ for a $\theta=15^\circ$ concave bend. 
For clarity only the values around the minimum are colored.
The lines near the bottom show the cylindrical segment ($\kappa=1/2\sigma$ for these plots) (red) and the straight
planes (black) used to model the bent graphene. All distances are scaled by $\sigma$.
Bottom: Same as middle for a convex bend.}
\label{fig:2dpot}
\end{figure}

{\em{Model calculation.}}\ 
We use a Lennard-Jones potential to model the interaction between a single cluster and the graphene
surface. The cluster is far enough away such that the fine detail of both the
cluster and the graphene sheet can be approximated by a solid sphere and continuous sheet, respectively.
We can then write the interaction using a standard Lennard-Jones potential \cite{jlennard-jones31},
scaled to make it dimensionless,

\begin{equation}
\frac{V(\vec{R})}{4\epsilon} = \frac{1}{A} \int d^2r\ \frac{\sigma^{12}}{|\vec{r} - \vec{R}|^{12}} -
                                                       \frac{\sigma^6}{|\vec{r} - \vec{R}|^6} 
\end{equation}

where $\epsilon$ is the interaction energy, $\sigma$ is the interaction distance parameter,
$\vec{R}$ denotes the cluster position and $\vec{r}$ is integrated over the infinite graphene sheet.
To more properly model a cluster, we integrated the above expression over the spherical volume as well.
However, the resulting potential obtained is qualitatively no different from a Lennard-Jones potential
with different parameters, so that step is neglected and we shall consider the cluster as a single entity at
the sphere's origin.
All distances can simply be scaled by $\sigma$, so the absolute value does not give quantitatively
different results.
We used this model to study the interaction between the cluster and a pleat in the graphene sheet. The pleat
was modeled by two straight sheets at the pleat angle $\theta$, inscribed by a cylinder segment of varying curvature $\kappa$.
to soften the sharp angle (see Fig. \ref{fig:2dpot}).

\begin{figure}[ht]
\includegraphics[width=2.7in, angle=270]{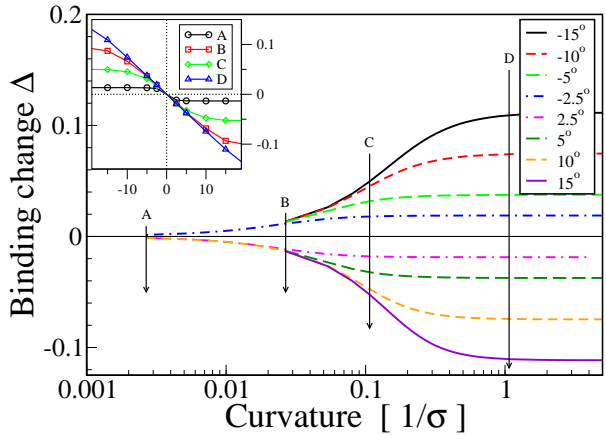}
\caption{(Color online) Semilog plot of $\Delta$ (see text) as a function of cylinder curvature.
The pleat angle $\theta$ increases monotonically from $-15^\circ$ to $15^\circ$, top to bottom.
Inset: Constant-curvature cuts as a function of $\theta$.}
\label{fig:curvplot}
\end{figure}

Figure \ref{fig:2dpot} shows the potential for a $15^\circ$ bend, inscribed with a cylinder of $\kappa=1/{2^{1/6}\sigma}$.
The concave bend shows a minimum relative to the flat pieces at the edge of the plot, which corresponds to
stronger binding. The minimum occurs due to increased surface area for interaction as clusters approach the pleat. For the convex
bend we observe the opposite, a decrease in binding due to decreased nearby surface area. To make this more quantitative, we
considered the binding energy relative to the flat surface as a function of curvature. We define the relative binding change
$\Delta = E_{curve}/E_{flat}-1$, where $E_{flat}$ is the energy minimum found for a flat surface.
Figure \ref{fig:curvplot} shows $\Delta$ as a function of the cylinder curvature $\kappa$.

For very low curvature (large cylinder radius), the binding energy of both concave and convex bends approach the
flat result. In the limit of zero curvature, they are identical.
In the limit of large curvature (small radius),
there is strong enhancement of binding for $\theta > 0$, and suppression for $\theta < 0$. The inset shows that, as a
function of $\theta$, there is a binding energy change for a wide range of curvatures. Furthermore, it is approximately symmetric
about the origin - the enhancement corresponding to a concave bend is roughly equal to the suppression corresponding
to a convex one. For the ranges of angles observed in experiment, there enhancement/suppression ranges from 2\% to over 10\%.\\
{\em{Discussion.}}\ 
The calculations show that typical bends, as found in experiments, can either enhance or suppress the total interaction by up to 15$\%$ compared to a flat surface. Our DFT calculations indicate a total binding of around 1 eV for $Ag_{20}$ on the planar graphitic surfaces, which is comparable to previous theoretical studies\cite{gwang03,dduffy98}. 

We use this value as a rough lower limit for bigger clusters like Ag$_{500}$, as used in our experiment.
Based on this, the calculations show that there is a minimum bend angle required to trap the clusters in the bend. Energetically,
the trapping starts between 2\% and 4\% enhanced binding (at room temperature). This corresponds to a bend angle between
3$^\circ$ and 5$^\circ$, as long the radius of curvature of the bend does not exceed 10 times the cluster radius,
or about 15 nm for Ag$_{500}$.
Similarly, convex bends with corresponding parameters cause effective barriers that cannot be traversed by cluster diffusion.\\

The results of our model calculation explain the experimental observation shown in Fig. \ref{fig:expt}. When clusters moving on the graphene surface encounter a concave bend of a pleat, they are trapped by the enhanced binding (potential well relative to the flat surface). They now have to diffuse in the potential well along the bend and aggregate to linear islands. Clusters landing on top of a pleat that has a flat surface are unable to overcome the barrier at the convex bends, and are trapped on top of the pleat where they have to aggregate.\\
By contrast, clusters moving on graphitic surfaces with an overall small angle of curvature do not "see" the curvature and aggregate as shown on Fig. \ref{fig:expt}(d).
Our DFT calculations show that the bonding of single silver atoms is oppositely influenced by the surface curvature.
Consequently, the manner in which bends in graphite affect atoms and clusters is qualitatively different.
The DFT calculations further show that silver dimers already behave classically and cluster-like. As the cluster size is increased further, only the total binding, and therefore the effective wells or barriers, caused by bends in graphitic pleats increases. For clusters with 500 silver atoms the potential wells and barriers are high enough to be used as effective guides or traps for cluster diffusion and aggregation, opening new routes for controlling anisotropic diffusion at nanometer scale.
\begin{acknowledgments}
This work was supported by DOE grants DE-FG02-02ER45995 and DE-FG02-97ER45660.
We would like to thank John J. Rehr for stimulating discussions.
\end{acknowledgments}
\vspace*{-5 mm}
\bibliography{master}
\end{document}